\documentclass[10pt,letterpaper,compsoc,conference]{iiswc22}

\usepackage{cite}
\usepackage{amsmath,amssymb,amsfonts}
\usepackage{algorithmic}
\usepackage{graphicx}
\usepackage[dvipsnames]{xcolor}
\usepackage[final]{microtype}
\usepackage[italic]{mathastext}
\usepackage{libertine}
\usepackage[T1]{fontenc}
\usepackage{textcomp}
\usepackage[varqu,varl]{zi4}
\usepackage[all]{nowidow}
\usepackage[auth-lg,affil-it]{authblk}
\usepackage[keeplastbox]{flushend}

\usepackage{subcaption}
\usepackage{tcolorbox}
\usepackage{xcolor}
\usepackage{listings}
\usepackage[hyphens]{url}

\begin{document}


\title{Performance Characterization of AutoNUMA Memory Tiering on Graph Analytics}


\renewcommand\Authsep{\qquad}
\renewcommand\Authand{\qquad}
\renewcommand\Authands{\qquad}


\author[1]{Diego Moura}
\author[2]{Daniel Moss\'e}
\author[3]{Vinicius Petrucci}
\affil[1]{Federal University of Bahia, Brazil}
\affil[2]{University of Pittsburgh, USA}
\affil[3]{Micron Technology, USA}

\maketitle


\begin{abstract}
Non-Volatile Memory (NVM) can deliver higher density and lower cost per bit when compared with DRAM. Its main drawback is that it is slower than DRAM. On the other hand, DRAM has scalability problems due to its cost and energy consumption. NVM will likely coexist with DRAM in computer systems and the biggest challenge is to know which data to allocate on each type of memory. A state-of-the-art approach is AutoNUMA, in the Linux kernel. Prior work is limited to measuring AutoNUMA solely in terms of the application execution time, without understanding AutoNUMA's behavior. In this work we provide a more in-depth characterization of AutoNUMA, for instance, identifying where exactly a set of pages are allocated, while keeping track of promotion and demotion decisions performed by AutoNUMA. Our analysis shows that AutoNUMA's benefits can be modest when running graph processing applications, or \textit{graph analytics}, because most pages have only one access over the entire execution time and other pages accesses have no temporal locality. We make a case for exploring application characteristics using object-level mappings between DRAM and NVM. Our preliminary experiments show that an object-level memory tiering can better capture the application behavior and reduce the execution time of graph analytics by 21\% (avg) and 51\% (max), when compared to AutoNUMA, while significantly reducing the number of memory accesses in NVM.
\end{abstract}

\section{Introduction}

The second millennium has brought about large transformations in the way we rely on computing systems to solve complex problems. Connections are everywhere: in social networks, in molecules, in biology, etc. With these applications, we are witnessing a real data explosion and the trend is for this volume to grow significantly over time \cite{YAQOOB20161231}. As we have more data, the demand increases significantly for memory. 

Because of the importance of analyzing graph applications (e.g., social networks, maps, etc), the GAPBS \cite{gapbs} benchmark suite has become very popular. These applications work by analyzing very large datasets in a graph format, using data points as nodes and relationships as edges. In addition to working with large inputs, this type of application is characterized by having poor temporal and spatial locality, which makes the memory subsystem a performance bottleneck \cite{Ozdal2016}.

To address the memory capacity problem, we could try to scale the capacity of DRAM. However, this is hard to accomplish due to its power, cost, and physical limitations (density problem). Non-volatile memory (NVM), also known as Persistent Memory, is a viable alternative because its high density, low standby power and low cost per bit. Although these are common features in most NVM products, there are other features that distinguish one NVM from another, such as granularity of access, the communication channel with the processor, the latency  ratio between load and store, etc. 

Heterogeneous memories, like DRAM+NVM, are likely to become common, similarly to heterogeneous processors (e.g.,  big.LITTLE\cite{Arm2013}), which allows for balancing performance and cost. This research focuses on systems with two types of memory, one with lower capacity and faster (DRAM), and another with higher capacity and slower (NVM).
In particular, we  explore a specific NVM incarnation, the Optane Persistent Memory Module \cite{inteloptane}, a Non-Volatile DIMM (NVDIMM)\footnote{Henceforth, Optane, NVDIMM, and NVM are used interchangeably.} technology that uses the DRAM DIMM form factor and is directly plug-compatible with system memory. 

As the latency costs of accessing pages on NVM can be much higher than DRAM, allocating frequently accessed (hot) pages on DRAM is essential for minimizing the performance impact on applications. A recent proposed memory tiering solution is Intel's AutoNUMA \cite{AutoNUMA}, for the Linux kernel. The few works in the literature that study the AutoNUMA performance only measured execution time. To our knowledge, no prior investigation was carried out to track and understand the decisions of AutoNUMA for memory tiering.

In this paper we characterize well-known graph applications using Intel's PEBS (Processor Event-Based Sampling) memory profiling mechanism, provided by the \verb@perf-mem@ tool on Linux. We use the \verb@perf-mem@ tool to collect memory samples (not trace) of memory access (both loads and stores) relying on the \verb@perf_events@ kernel interface. From the samples we identified that approximately 
60\% of pages (on average) accessed out of cache  experience only one touch per page. This is somewhat expected since graph analytics has irregular access patterns. Considering that AutoNUMA performs migrations of hot pages to fast memory (tier-1) based on the interval between at least two accesses to the same page, graph workloads can inhibit AutoNUMA  from making effective memory management decisions. 

For AutoNUMA to work effectively, there must be non-uniform number of accesses to the pages, that is, some pages with a high number of accesses and others with low number of accesses. In this way it is possible to distinguish cold from hot pages. Another way to address the problem is to prefetch the pages, that is, to anticipate which pages are hot, even before they are used for the first time. But for that it is necessary to have a predictable pattern in the page accesses. The graph applications that we are studying do not show this behavior. However, tracking memory access, when combined with information about application allocations, can provide valuable information. This is because allocations from applications are typically done by objects, not pages. In this way, we can look at the problem of knowing which \textit{address range} has the highest chance of the next memory access, rather than the problem from predicting which \textit{page} will be accessed next.

This paper makes the following contributions:

\begin{itemize}
    \item We present the characterization of irregular real-world memory access from graph-based applications using DRAM+NVM. Our characterization results show that the vast majority of pages are accessed only once.
    \item We quantify the decisions performed by AutoNUMA running graph applications with irregular memory access on tiered memory systems. The results show that AutoNUMA is not able to effectively identify pages for promotion to DRAM.
    \item We propose and quantify a new idea of using object-level mapping rather than page-level management. Our preliminary results, using a static algorithm, on average 21\% and up to 51\% reduction in execution time compared to AutoNUMA.
\end{itemize}

\section{Background}

\subsection{Non-Volatile Memory}

In our implementation and experiments, we use Intel's
Optane DC Persistent Memory Module (or just ``Optane"), which is a cost-effective technology that comes closest to DRAM concerning performance. Furthermore, because these NVMs are packaged in DIMMS and are on the same bus/channel of the DRAM, it is not possible to achieve the theoretical maximum bandwidth for DRAM and NVM when used together. Through this common bus, the CPU connects each of its two integrated memory controllers (iMC) to three channels. 

In addition to the NVM capacity (up to 3TB), three other features should be highlighted: latency, bandwidth, and asymmetry between read and write operations. The read latency of the Optane compared to DRAM is about 3x for random accesses and 2x for sequential access \cite{joseph2019}. The reason is that  adjacent requests are combined in a buffer before a store happens using the 256-byte internal granularity. This is a signal that if we coordinate the access to the NVM, we can explore  locality, which can reduce latency and write-amplification. The store latency is difficult to measure because there is no mechanism to record when a store physically reaches the NVM (due to write caches and delayed writes). 

When comparing Optane's bandwidth with DRAM for a different number of threads (between 1 and 21 threads), prior work \cite{joseph2019} has shown that as the number of threads increases, the difference between the two technologies increases. For reading, DRAM achieves values above 100 GB/s while Optane achieves about 40 GB/s. For writing, the difference is even greater, about 80 GB/s for DRAM and around 14 GB/s for Optane when using 21 threads. The authors also showed that the read latency for random load operations is 3x slower on Optane vs DRAM and 2x when data is accessed in a sequential pattern.
Other researchers have highlighted \cite{Jian2020} that there are nuances when using this new technology, most notably (a) stores cost more (has higher latency) than loads, (b) accesses with request sizes less than 256 bytes tends to waste bandwidth, and (c) writing results in write amplification. 

\begin{figure}[t]
    \centering
    \includegraphics[scale=0.75]{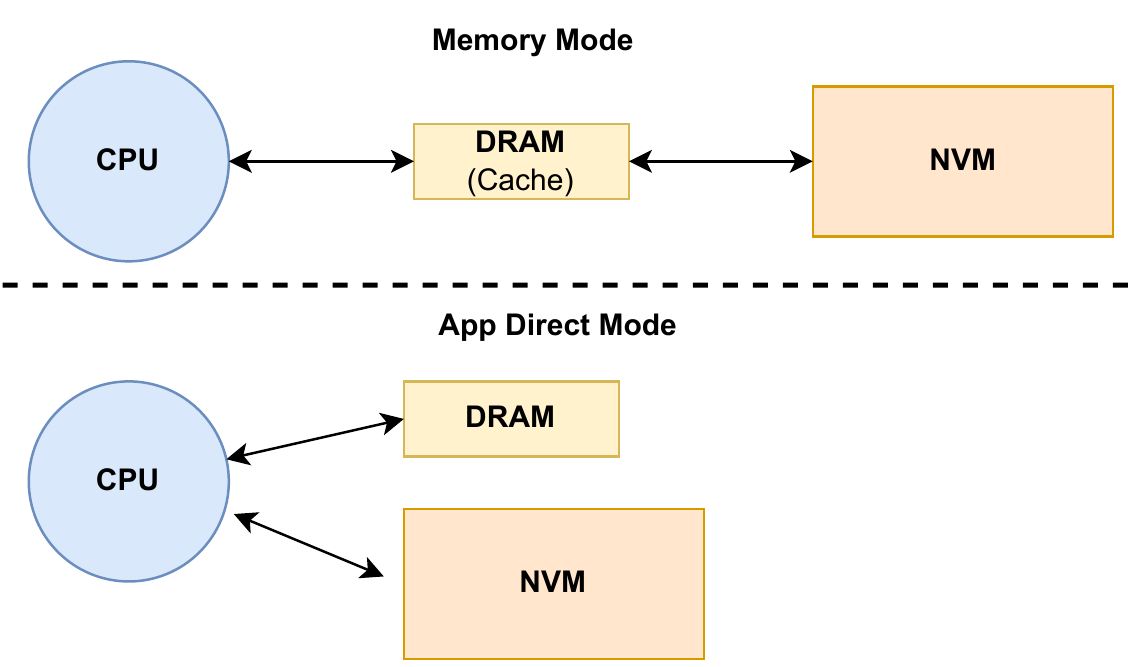}
    \caption{Usage of Non-Volatile Memory in a system: Memory Mode versus App Direct Mode.}
    \label{fig:memory_mode_vs_app_direct}
\end{figure}

The Optane has two operation modes: \textit{Memory Mode} and \textit{App Direct Mode} as Figure \ref{fig:memory_mode_vs_app_direct} shows.
In \textit{Memory Mode}, the DRAM works as a cache for the Optane memory. All memory hierarchy  management is made similar to using only DRAM, without any changes to the application. Note that this setup will have little impact on the performance of applications with a \textit{small memory footprint} (smaller than the size of the DRAM). Besides, the Optane in this mode does not work as a persistent storage device but as a large memory. In the other mode of operation, called \textit{App Direct Mode}, we can use the NVM as peer memory and perform data persistence. However, unlike \textit{Memory Mode}, the user explicitly defines which part of the application will use NVM.
 
Because \textit{Memory Mode} does not allow the control of which data will be allocated on DRAM or NVM, we focus on \textit{App Direct Mode}. Our work uses a variant of \textit{App Direct Mode}, supported by the Linux kernel version 5.1, that makes it possible to set up the NVM as a separate memory NUMA (Non-Uniform Memory Access) node \cite{PMEM_numanode}. In this configuration, NVM does not use persistence and data is volatile just like DRAM. Through memory \textit{affinity} command, it is possible to decide on what type of memory the allocations will be made. 

\subsection{Memory Tiering with AutoNUMA}

NUMA (Non-uniform memory access) machines now include CPUs with attached DRAM and CPU-less memory node (e.g., NVM as a memory NUMA node). This setup is also called a \textit{memory tiering} system, because the performance of the different types of memory are
usually different (e.g., DRAM as tier-1 and NVM as tier-2). The original AutoNUMA algorithm (without considering memory tiering) identifies pages recently accessed by tasks on a local CPU from a remote NUMA node and migrates these pages to local NUMA node, trying to minimize remote access latency. With \textit{memory tiering}, the goal is to manage data between two types of memory, for example, DRAM (fast memory) and NVM (slow memory), considering all memory requests coming from tasks running on the same CPU.

Leveraging the original AutoNUMA infrastructure, a new patch series has been proposed by Intel for the memory tiering scenario \cite{AutoNUMA}. This new AutoNUMA version is able to identify cold pages in DRAM and   migrate them to NVM (called \textit{demotion}), and to identify hot pages in NVM and  migrate them to DRAM (\textit{promotion}). The methodology used by AutoNUMA to mark a candidate page as \textit{hot} is based on the existing NUMA balancing page table scanning and hint page fault mechanisms. In this process, a set of pages is chosen to be scanned and the status of these pages is changed to \verb@PROT_NONE@. This causes a new access to such pages to trigger a hint page fault. A timestamp called \textit{scan time} is recorded during the page scan process. When a marked page is accessed, a hint page fault  occurs, and a timestamp called \textit{hint page fault time} is recorded. Then, the difference between page fault time and scan time is calculated, which is called \textit{hint page fault latency}. The intuition is that the shorter the hint page fault latency of a page, the higher its access frequency is likely to be.

AutoNUMA specifies a threshold for the hint page fault latency to classify a page as hot or cold. As it is difficult to determine a threshold in practice, AutoNUMA adjusts this threshold  dynamically, based on two new concepts: \textit{candidate promotion pages} and \textit{promotion rate limit}. Candidate promotion pages are pages that are below the threshold and may or may not be promoted. The promotion rate limit is a limit set statically by the system administrator. The default value is 35MB. The idea is that if the number of candidate promotion pages in a given interval is much more than the promotion rate limit, the threshold will be decreased to reduce the number of candidate promotion pages. Otherwise, the threshold will be increased to enlarge the number of candidate promotion pages. This technique is only triggered when there is no more space in the fast memory node (e.g., DRAM). If there is enough free space in the fast memory, the  threshold will not be used, and all pages can be promoted upon a hint page fault. This is a strategy to avoid the potentially long time to detect hot pages.

\section{Characterization Methodology}
\label{sec:methodology}

\begin{figure}[t]
    \centering 
    
    \begin{subfigure}{0.50\textwidth}
      \includegraphics[width=\linewidth]{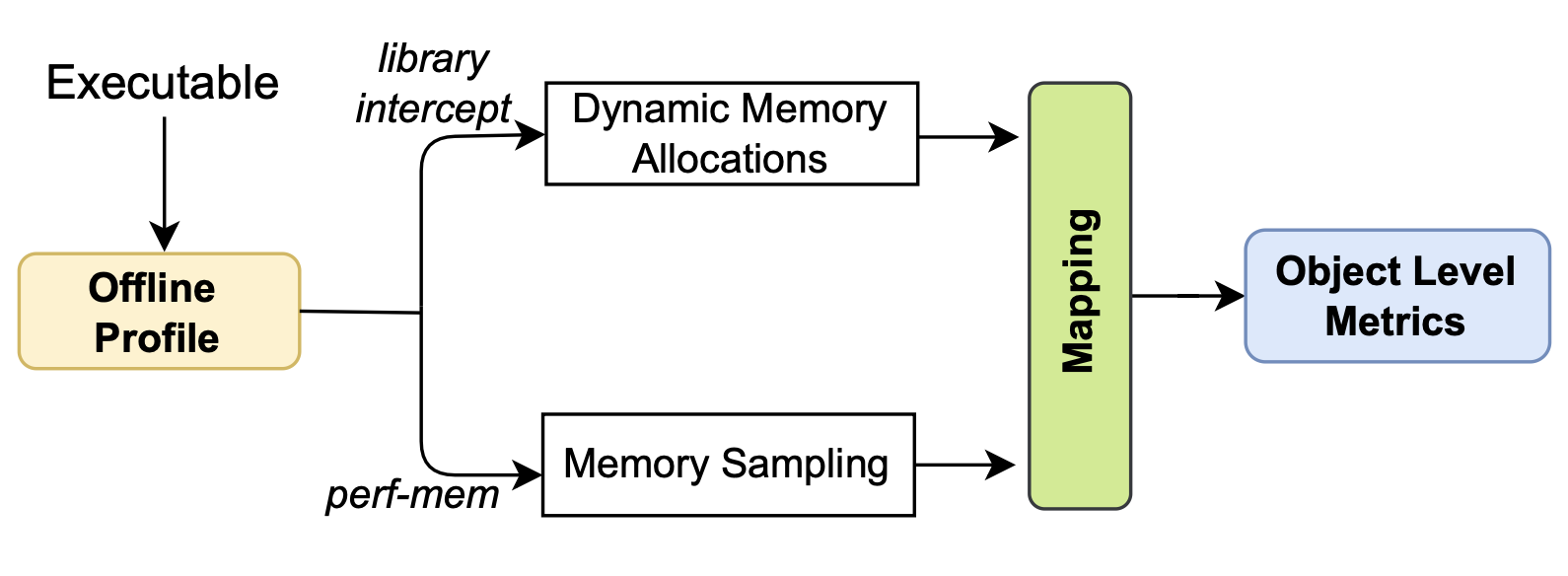}
    \end{subfigure}\hfil 
    
    \caption{Profiling workflow for memory object characterization.}
    \label{fig:methodology}
\end{figure}

We introduce a methodology for mapping memory samples to memory allocations as shown in Figure \ref{fig:methodology}. This methodology allows us to monitor application's memory accesses and aggregate those accesses to memory allocations over time. We use \verb@perf-mem@ tool to perform memory access sampling (see Section~\ref{sec:recording-memory}), while we keep track of dynamic memory allocations  (see Section~\ref{sec:tracking-memory}). 
We then associate each memory sample to a particular memory object allocated by the applications (see Section~\ref{sec:mapping-samples-to-objects}).

\subsection{Recording Memory Samples}
\label{sec:recording-memory}

In Linux, \verb@perf@ is a standard profiling infrastructure used to collect and analyze application performance and tracing events. We use \verb@perf-mem@ to sample (not trace) memory accesses (load and store instructions). 
A memory load \textit{sample} contains (a) the hierarchical level of memory where the access took place (i.e., DRAM, PMEM/NVM, L3, L2, L1 and LFB.),  (b) the corresponding memory address, which is used to calculate the memory object that this address belongs to, and (c) the latency cost in cycles to access the memory.  Note that ``store'' events can only be analyzed at the L1 level. We focus on load samples that occur outside caches (either on DRAM or NVM).

\subsection{Tracking Memory Allocations}
\label{sec:tracking-memory}
In addition to collecting memory samples, we also track the dynamic allocations performed by the applications.
When an application
needs to allocate memory blocks that are larger than a threshold (in Linux, this is given
by \verb@MMAP_THRESHOLD@ of 128 kilobytes), the \verb@malloc@ function allocates space in the MMS (Memory Mapping Segment) region using the system call \verb@mmap@, rather than using the \verb@sbrk@ system call to expand the Heap. Because our applications work with large memory allocations (multi-page granularity), we track memory allocations originating from the \verb@mmap@ system calls. 
We track memory allocations/deallocations with the \verb@syscall_intercept@ shared library used for hooking Linux system calls in userspace \cite{Rudoff17}.
When intercepting each allocation, we record the timestamp, the size of the allocation, the starting memory address, and the call stack.

\subsection{Mapping Memory Samples to Objects}
\label{sec:mapping-samples-to-objects}

We define a ``memory object'' as a contiguous memory region originating from a ``mmap'' syscall called by the C standard library or applications. Thus, the terms ``object'' and ``memory allocations'' are used interchangeably in our work. 

We use applications that make use of large chunks of memory (granularity of several pages). That's why we track \verb@mmap@ instead of \verb@malloc@. Since for each \verb@mmap@ intercept there is a memory address range and each sample has a memory address associated with it, we can perform the mapping between memory samples and \textit{objects}). This is done after recording the address and timestamp of each memory sample and associating this information with a particular object based on the call stack. This lets us calculate metrics at object-level (i.e., for each object), such as access frequency and lifetime, and decide which objects to allocate on DRAM and which to allocate on NVM. 

In summary, using memory traces collected through the Linux \verb@perf-mem@ tool and identifying dynamic allocation blocks through \verb@mmap@ system call intercepting, we are able to sample memory accesses and associate such samples with application-level memory objects. As we will show next in our experiments, this approach can bring new opportunities to improve application memory management for performance.

\section{Experimental Setup}
\label{sec:experimental_setup}

\subsection{Benchmark Description}

We use GAPBS \cite{gapbs} applications that process big data in a graph format, considering data points as nodes and relationships as edges. The applications used are: \textit{Betweenness Centrality} (BC), \textit{Breadth First Search} (BFS) and \textit{Connected Components} (CC). BC is a graph application that measures the importance of a node in a graph. In social networks analysis, it is actively used for computing the user ``influence'' index. The betweenness centrality for each vertex is the number of shortest paths that pass through the vertex. BFS is an application responsible for traversing a graph from the root node (or some arbitrary node of a graph) and exploring all the adjacent nodes (nodes which are directly connected to root node). BFS can be used in different scenarios such as to locate all the nearest or adjacent nodes in a peer-to-peer network. CC is an application that identifies the maximum sets of vertices reachable from each other in an undirected graph. As input we use the datasets \verb@kron@ and \verb@urand@ generated using the GAPBS suite \cite{gapbs}. We selected those datasets from the GAPBS suite exclusively because the other datasets (``twitter", ``road", and ``web") had memory footprints too small to be used in our system. Therefore, we have experimented with 6 application-dataset combinations.

\subsection{System Hardware and Software}
Our experiments run on a real dual-socket machine, each with Intel(R) Xeon(R) Gold 6240 CPU @ 2.60GHz, 18 cores, with hybrid memory: 192 GB of DRAM (6x32GB DIMMs) and 768 GB of NVM (6x128GB DIMMs, in this case, Intel Optane Memory) per socket. To eliminate the influence of non-uniform memory access,  all experiments use only one socket. In addition, hyperthreading is disabled and the governor is configured as performance (i.e., CPU frequency is set at the maximum allowed).
We use the latest version of AutoNUMA called Tiering v0.8 \cite{AutoNUMA} running on Fedora 32 with Linux kernel 5.15. To bring up NVM as a new NUMA node, we used \verb@ndctl@ and \verb@daxctl@ tools version 71.1.

\section{Characterization of Graph Applications}
\label{sec:graph_application_caracterization}


\subsection{Page Access Distribution}

The distribution of accesses on different memory hierarchy levels can tell us, for example, if the application benefits or not from the use of caches. By identifying what percentage of memory accesses occur outside caches (either in DRAM or NVM), we can characterize application's data access locality.

\begin{figure}[t]
    \centering 
    
    \begin{subfigure}{0.47\textwidth}
      \includegraphics[width=\linewidth]{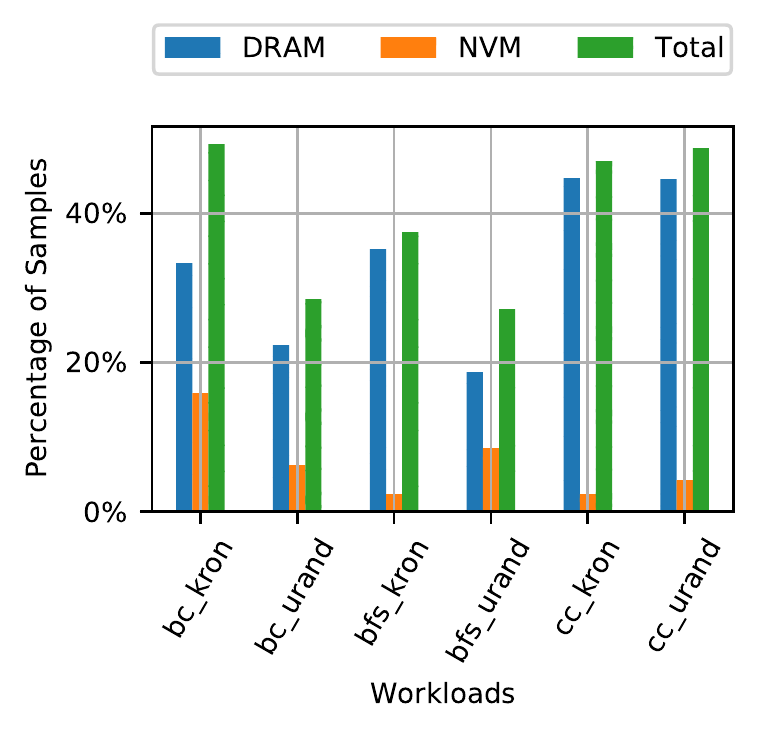}
      \label{fig:4}
    \end{subfigure}\hfil 
    
    \caption{Percentage of memory samples mapped to DRAM and NVM for different graph applications/datasets.}
    \label{fig:samples_distribution_mem_level}
\end{figure}

Figure \ref{fig:samples_distribution_mem_level} shows the percentage of total memory samples experienced outside caches when AutoNUMA is enabled in the system. For these applications, at least 25\% of the total (green bar in the figure) samples occur externally (DRAM + NVM), and this number can reach close to 50\%. This pattern is typical of applications with irregular/random access like graph analytics, which makes it difficult to explore existing caches. 
Since the latency of accessing data on NVM is much higher than on DRAM, it is paramount for memory tiering solutions to  minimize the accesses that occur in the NVM by effectively moving pages from the NVM to the DRAM. As we will discuss in Section~\ref{sec:AutoNUMA_track_decisions}, this can be done \textit{reactively} as in AutoNUMA, which identifies candidate pages most likely to be accessed based on past accesses, or \textit{proactively} by trying to anticipate what pages will be accessed in the future. 

\subsection{Access Frequency per Page}
\label{sec:access_freq_page}

When pages are assigned to NVM (tier-2), page reuse is an important metric as it gives us a guidance of which page to migrate from NVM to the DRAM. For instance, if reuse is low for a page, it is unlikely the page will be reused in the near future, thus the cost of migration of the page to DRAM may outweigh the benefit. 

\begin{figure}[t]
    \centering 
    
    \begin{subfigure}{0.47\textwidth}
      \includegraphics[width=\linewidth]{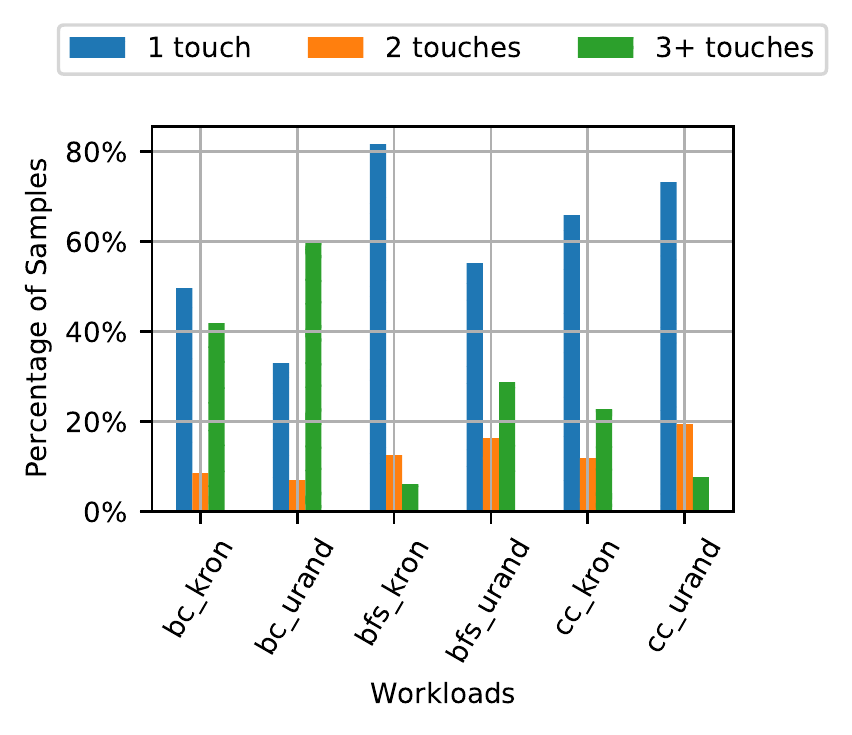}
      \label{fig:4}
    \end{subfigure}\hfil 
    
    \caption{Percentage of page accesses with 1, 2, or 3+ touches for different applications/datasets (each access occurred external to caches).}
    \label{fig:frequency_of_pages_access}
\end{figure}

Figure \ref{fig:frequency_of_pages_access} shows the number of touches to the pages accessed outside the cache (DRAM and NVM). 
The blue (first) bar shows that at least 33\% of all external accesses touch a page only once (\verb@bc_urand@), reaching up to 80\% (\verb@bfs_kron@). This means a \textit{reactive} approach as used by AutoNUMA cannot characterize as hot at least half of pages accessed outside the cache, except for \verb@bc_urand@ workload. 
Focusing on the pages with only two touches, the percentage of memory accesses reduces to about 10\% on average. Looking at these pages with only two touches, if they are considered hot by  AutoNUMA after the second touch, migrating them to DRAM will not improve performance (but it will use more DRAM space and bandwidth), since no new accesses will occur on this page. We note that this analysis considers the entire execution time of the application (not a particualr phase/interval) and relies on memory sampling (not tracing) as an approximation of the memory accesses. Performing tracing of all memory accesses to support memory tiering at runtime is not practical given the size of the application.

\begin{figure}[htbp]
    \centering 
    \begin{subfigure}{0.24\textwidth}
      \includegraphics[width=\linewidth]{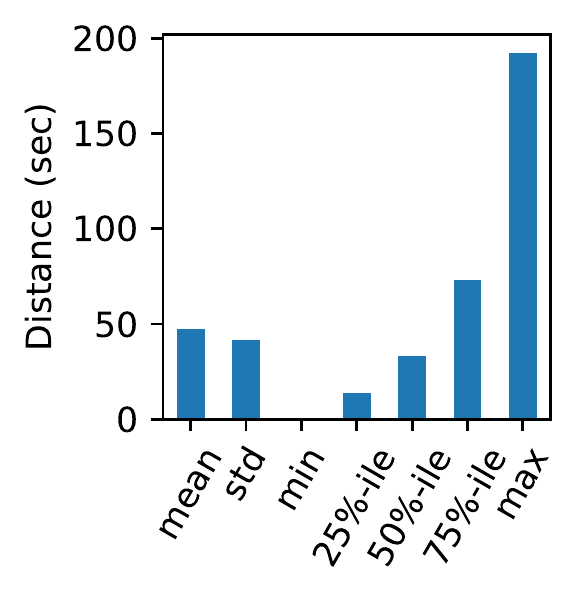}
      \caption{bc\_kron.}
      \label{fig:1}
    \end{subfigure}\hfil 
    \begin{subfigure}{0.24\textwidth}
      \includegraphics[width=\linewidth]{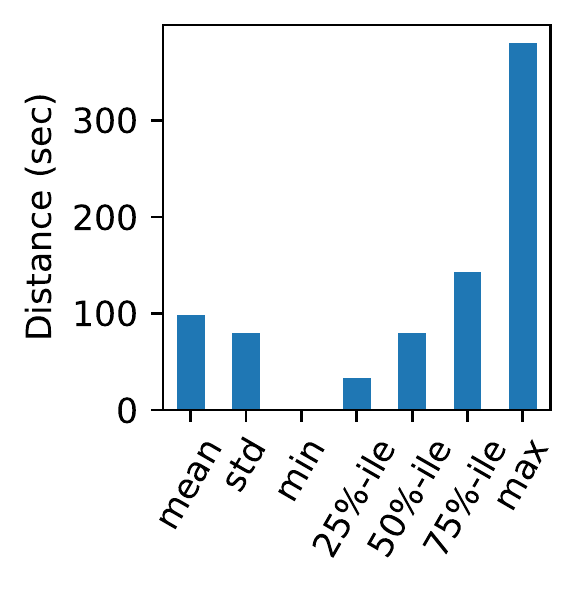}
      \caption{bc\_urand.}
      \label{fig:2}
    \end{subfigure}\hfil 
    \begin{subfigure}{0.24\textwidth}
      \includegraphics[width=\linewidth]{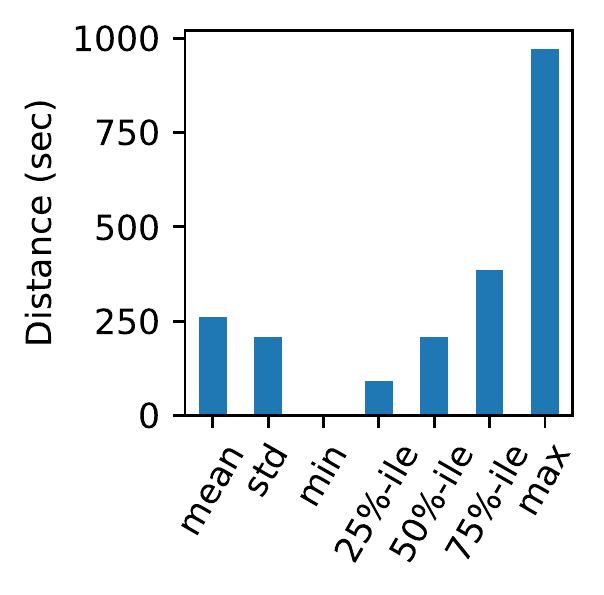}
      \caption{bfs\_kron.}
      \label{fig:1}
    \end{subfigure}\hfil 
    \begin{subfigure}{0.24\textwidth}
      \includegraphics[width=\linewidth]{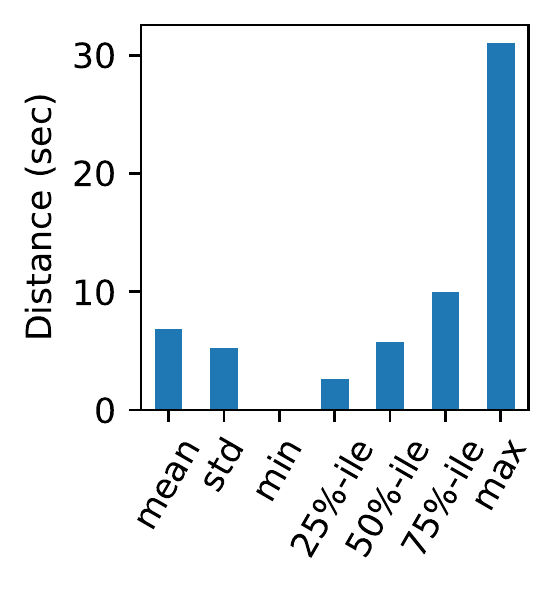}
      \caption{bfs\_urand.}
      \label{fig:2}
    \end{subfigure}\hfil 
     \begin{subfigure}{0.24\textwidth}
      \includegraphics[width=\linewidth]{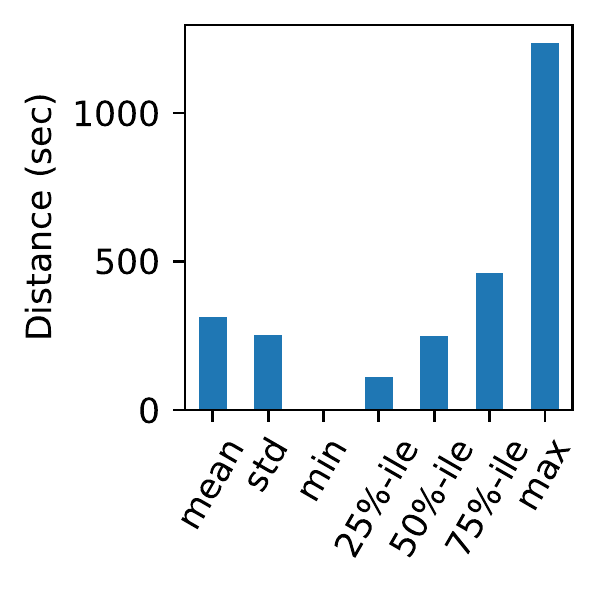}
      \caption{cc\_kron.}
      \label{fig:1}
    \end{subfigure}\hfil 
    \begin{subfigure}{0.24\textwidth}
      \includegraphics[width=\linewidth]{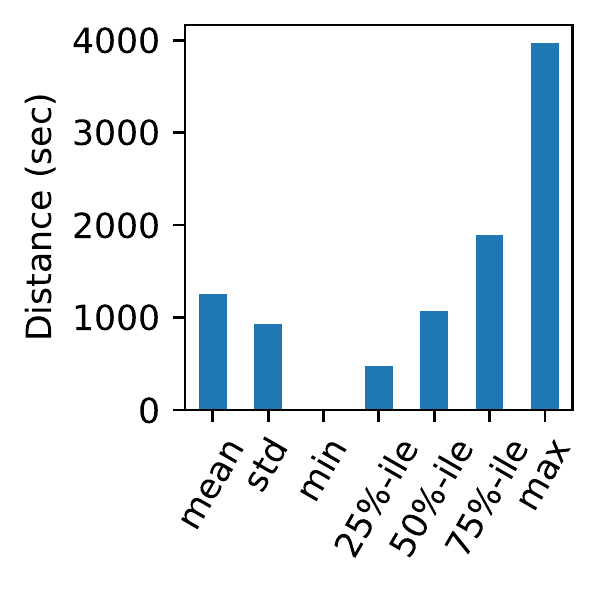}
      \caption{cc\_urand.}
      \label{fig:2}
    \end{subfigure}\hfil 
    \caption{Statistics of page reuse in time. We consider pages touched exactly twice and associated with the most accessed memory object allocated on NVM.}
    \label{fig:two_access_distance}
\end{figure}

The results shown in Figure \ref{fig:frequency_of_pages_access} do not distinguish which pages are allocated on DRAM or NVM. 
Therefore, we perform further experiments characterizing each application object with most accesses on NVM; that is, the hottest object of each application. To simplify our analysis, we extract all pages of such objects having only two accesses on NVM, since a page can be classified by AutoNUMA with at least two accesses. Based on this experiment, we would like to answer the following questions: (1) \textit{Is the interval between these two accesses regular?} (2) \textit{Can a page accessed at least twice on NVM be found in DRAM}? The first question can give us an idea of the difficulty of estimating if a page is hot. The second question can tell us if there are pages being promoted from NVM to DRAM given at least two touches. 

We first analyze the time between memory accesses for pages that had exactly two access (see Figure \ref{fig:two_access_distance}). This was calculated using the difference between the timestamps of the two consecutive accesses (recorded samples) to the same page. The value is given in seconds. We report the minimum distance (min), the maximum distance (max), and average distance (average), as well as 25th, 50th, and 75th percentiles. We notice that applications have a wide dispersion of reuse over time, as the standard deviation is close to the mean. For example, in the case of \verb@bc_kron@ workload, in 25\% (25th percentile) of the cases the interval between the two (sampled) access was a maximum of 14 seconds. Looking between the max value and the 75th percentile value, there are 25\% of samples with the interval between two accesses above 73 seconds. Even if there was a dynamic mechanism to characterize when a page is hot, in this case it still may not work well because the pages have a very high variance in reuse over time. This makes the work performed by AutoNUMA for identifying hot pages  challenging.

We also analyze (but do not show in graphs) the percentage of 2-access pages that were promoted from NVM to DRAM. When a page is accessed on NVM and later accessed on DRAM, we consider that the page was promoted to DRAM (the timestamp of the page accessed on DRAM is higher than the timestamp of access on NVM). For all applications, at most 1.3\% of the pages analyzed were accessed on the NVM and later on the DRAM. 
Some applications had no page accessed in the NVM and then accessed in the DRAM. This is a consequence of the irregularity in the access interval between the pages, which makes it hard to characterize a page as hot and candidate for promotion. In particular, as a reactive approach, AutoNUMA relies on a few page touches over time to identify hot pages and may not perform well for the graph application scenario, as a large number of pages only experiences one or two touches. One may consider a scheme to anticipate future accesses  by promoting the entire object or by promoting a range of addresses within the object as a proactive approach.

\section{Characterization of AutoNUMA}
\label{sec:AutoNUMA_track_decisions}

To understand the behavior of AutoNUMA and track its memory management decisions, we experiment with
graph applications and datasets that have memory footprint (228-292GB) greater than the capacity of DRAM memory (192GB).

\subsection{Tracking Page Accesses on DRAM vs NVM}

Hot pages that are not found in the caches require frequent external memory access. Table \ref{tab:samples_outside_from_cache} shows the percentage of collected memory samples whose access occurred outside the caches, and where the accesses occurred. 

\begin{table}[h]
\centering
{\begin{tabular}{c|c|c|c}
\hline
\multicolumn{1}{c}{\bfseries Workload} &
\multicolumn{1}{c}{\bfseries Outside Cache} &
\multicolumn{1}{c}{\bfseries Pages in DRAM} &
\multicolumn{1}{c}{\bfseries Pages in NVM}  \\ \hline
bc\_kron & 49.1\% & 67.69\% & 32.31\% \\
bc\_urand & 28.5\% & 78.18\% & 21.82\%\\
bfs\_kron & 37.4\% & 93.87\% & 6.13\%\\
bfs\_urand & 27.1\% & 68.83\%  & 31.17\% \\
cc\_kron & 46.9\% & 95.08\% & 4.92\% \\
cc\_urand & 48.6\% & 91.48\% & 8.52\%  \\

\end{tabular}}
\caption{Percentage of page samples that hit on DRAM or NVM for different workloads.}
\label{tab:samples_outside_from_cache}
\end{table}

The second column of Table \ref{tab:samples_outside_from_cache} shows that a significant portion of the sampled memory accesses took place in DRAM or NVM (27\%-49\%). The third and fourth columns help us answer the following question: ``\textit{When the external access occurred, was the page allocated in DRAM or NVM}"? We note that the amount of NVM access is not dependent on the application nor on the dataset, but a combination thereof, making it more difficult to predict  such behavior.  Thus, deeper performance analysis of the application-dataset combination is needed. For example, the behavior depends on the system characteristics (e.g., available memory) and application-dataset behavior (e.g., sequence of memory allocations).

\begin{table}[htbp]
\centering
{\begin{tabular}{c|c|c}
\hline
\multicolumn{1}{c}{\bfseries Application} &
\multicolumn{1}{c}{\bfseries DRAM Access Cost} &
\multicolumn{1}{c}{\bfseries NVM Access Cost} \\ \hline

bc\_kron & 37.53\% & 62.47\% \\
bc\_urand & 62.95\% & 37.05\% \\
bfs\_kron & 79.81\% & 20.19\%  \\
bfs\_urand & 28.20\% & 71.80\%  \\
cc\_kron & 89.51\% & 10.49\% \\
cc\_urand & 80.30\% & 19.70\%   \\

\end{tabular}}
\caption{ Percentage of total latency (in cycles) from samples that hit on DRAM or NVM for different workloads (ordered by high to low NVM Access Cost).}
\label{tab:access_cost}
\end{table}

Table \ref{tab:access_cost} shows the cost of all accesses outside the cache, distributed between DRAM and NVM. This cost is calculated using the number of processor cycles as a reference. For instance, \verb@bfs_urand@ and \verb@bc_kron@ experienced nearly 1/3 of external accesses on NVM (see Table \ref{tab:samples_outside_from_cache}), but such accesses represent more than half of the cost of external accesses (see Table \ref{tab:access_cost}). This confirms the high costs of accessing pages in NVM. In the trace obtained by \verb@perf-mem@, for each memory access there is an associated cost.  To obtain the values of Table \ref{tab:access_cost}, we added all access costs that occurred in the DRAM and divided by the cost of all external accesses (DRAM + NVM). The same procedure was done for the calculation of access costs in the NVM. It is important to note that the costs of accessing both the DRAM and the NVM are not the same for all requests. Therefore, it is not possible to make a direct correlation from the values contained in Tables \ref{tab:samples_outside_from_cache} and \ref{tab:access_cost}. A fact that contributes to the difference in the cost of access is whether or not an access is preceded by a TLB miss, as explored below. 

\begin{table}[htbp]
\centering

\begin{tabular}{l|c|c|c|c}
\hline
\bfseries Application & \multicolumn{2}{c}{\bfseries DRAM} & \multicolumn{2}{c}{\bfseries NVM} \\
& TLB Hit & TLB Miss & TLB Hit & TLB Miss\\
\hline

bc\_kron & 659  & 772 & 1833 & 2727 \\
bc\_urand & 1675  & 1617 &  2862 & 3439 \\
bfs\_kron &  404 & 490 & 1572 & 2218 \\
bfs\_urand & 578 &  734 & 2632 & 4183\\
cc\_kron & 315 & 866 & 1170 & 2975\\
cc\_urand & 325 & 903 & 1345 & 4141 \\

\end{tabular}

\caption{Average cost (in number of cycles) of external cache accesses, grouped by TLB hit and TLB miss.}
\label{tab:tlb}
\end{table}

Table \ref{tab:tlb} shows the average cost of access in number of cycles of accesses that occur externally (DRAM or NVM). For each type of external access (if DRAM or NVM), we separated the accesses that came preceded by TLB hit and TLB miss. We clearly see that external accesses preceded by a TLB miss are more expensive. This is highlighted for example for \verb@bfs_urand@ which had more than 90\% of its NVM access preceded by a TLB miss. This cost is much worse if the memory access occurs in the NVM. For example, for \verb@cc_urand@ the average cost of accesses for NVM + TLB miss is about 12x more than the DRAM + TLB hit case. 

\begin{tcolorbox}
\textbf{Finding 1}: The accesses occurred in the NVM preceded by TLB have an average access cost of 4x, reaching up to 5.7x more compared to the access costs to DRAM preceded by TLB miss.
\end{tcolorbox}

\subsection{Relating Page Accesses to Objects}

\begin{figure}[t]
\centering 
    \begin{subfigure}{0.47\textwidth}
      \includegraphics[width=\linewidth]{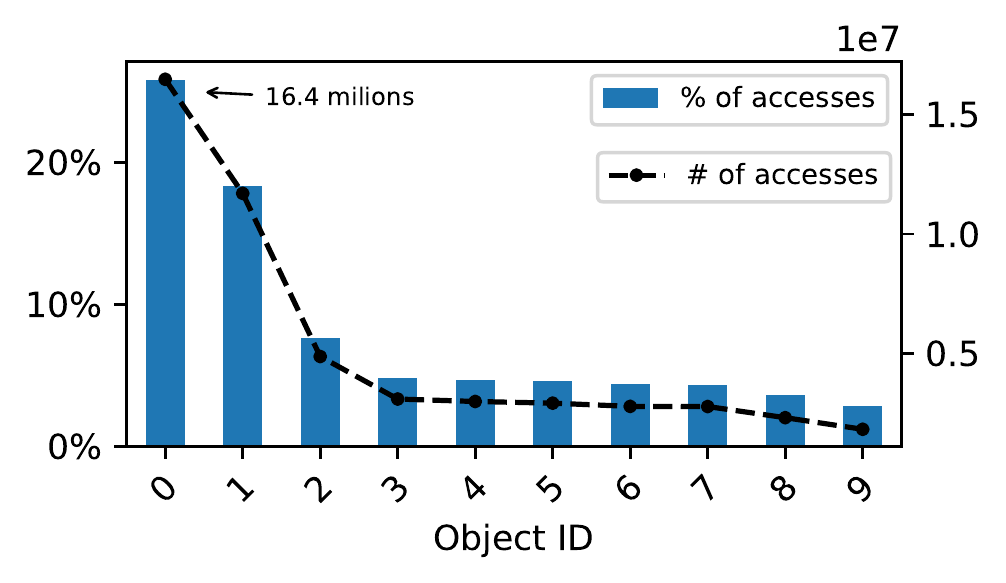}
      \caption{DRAM: Top 10 objects with most memory accesses.}
      \label{fig:2}
    \end{subfigure}\hfil 
    \begin{subfigure}{0.47\textwidth}
      \includegraphics[width=\linewidth]{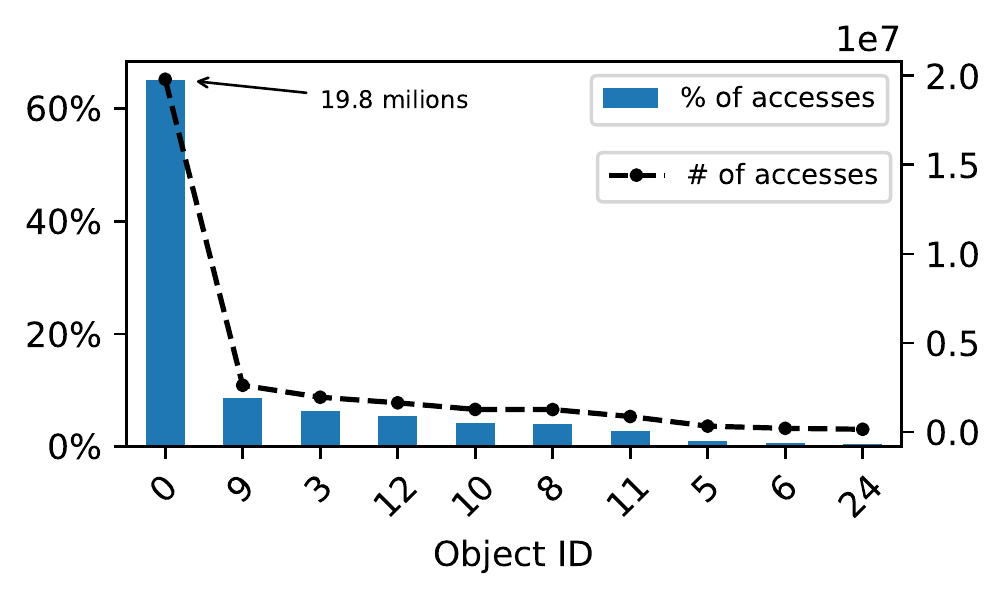}
      \caption{NVM: Top 10 objects with most memory accesses.}
      \label{fig:1}
    \end{subfigure}\hfil 
    \caption{Percentage of memory samples (DRAM at the top, NVM at the bottom) associated with specific objects (workload \emph{bc\_kron}).}    
\label{fig:page_access_to_objects}
\end{figure}

As discussed previously, graph applications may not have hot pages, but we would like to know if memory \textit{objects} (not individual \textit{pages}) concentrate most external accesses. Due to space limitations, from here on we will only detail the results of an illustrative  workload, namely \verb@bc_kron@. However, similar features are found in other workloads. Figure \ref{fig:page_access_to_objects}a and  \ref{fig:page_access_to_objects}b shows the proportional and absolute distribution of accesses in DRAM and NVM for the top 10 objects for \verb@bc_kron@, respectively.
The figure shows the object ID on the x-axis, the percentage of pages accessed on the left y-axis, and the absolute value of page accessed on the right y-axis.

We notice that 65\% of the memory accesses in the NVM are concentrated only in a single object (object 0). This is the same object that also received the highest number of accesses in the DRAM (object 0).

\begin{tcolorbox}
\textbf{Finding 2}: Very few objects concentrate the majority of memory accesses in the NVM. For instance, the workload \verb@bc_kron@ has a single object with about 60\% of all accesses in the NVM. This is even more pronounced in other workloads reaching up to 90\% as is the case with \verb@bfs_urand@ and \verb@cc_urand@. 
\end{tcolorbox}

After looking at the page access distribution across DRAM and NVM, a question that arises is: \emph{How could AutoNUMA allocate about half of the pages of this object in the DRAM and the other half in the NVM?} This is an interesting question because, as we have seen in Section \ref{sec:access_freq_page}, the pages of this application have a more uniform distribution of access. We will answer this question in the next section by collecting additional information.

\subsection{Object-level Analysis}

\begin{figure}[t]
    \centering 
     \includegraphics[scale=1.15]{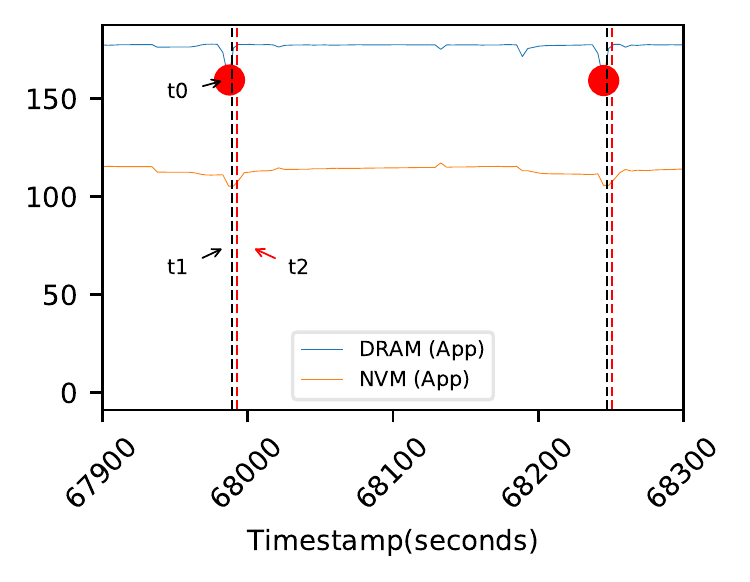}
    \caption{Identification in time for the allocations of objects with ID 0 and ID 9 (\emph{bc\_kron}) workload.}
    \label{fig:allocations_top_1}
\end{figure}

To understand in more detail why AutoNUMA allocated the hot object 0 on both DRAM and NVM, we tracked when object 0 was allocated, and how memory is allocated over time, as shown in Figure \ref{fig:allocations_top_1}. The x-axis refers to time and the y-axis to the current memory allocation amount in gigabytes (GB). The vertical red dashed line shows the moment (time \verb@t2@) at which object 0 was allocated. We notice that this allocation  occurs right after a memory release from DRAM (blue dot inside the red circle) at time \verb@t0@ (this behavior recurs over time--see time 68250 in the plot). The amount of memory released was approximately 13GB and object 0 has an allocation size of 8GB, which means this object could be fully allocated in DRAM at this moment.

However, another 8GB object (ID 1) is allocated before Object 0, see vertical black dashed line (time \verb@t1@). Figure \ref{fig:page_access_to_objects} shows that the object 1 only exists in the DRAM, which leads us to conclude that it was responsible for consuming most of the free space in the DRAM. 
Thus, pages of the object 0 were allocated in the DRAM not because they were characterized as hot, but because there was a space recently freed by another object.

\begin{tcolorbox}
\textbf{Finding 3}:  Pages are not only allocated in DRAM  because they are characterized as hot by AutoNUMA, but also because the AutoNUMA policy always allocates in DRAM first as long as there is space available.
\end{tcolorbox}

\subsection{Memory Access Pattern}
\label{subsec:Memory_Access_Pattern}


%

\begin{figure}[htbp]
    \begin{subfigure}{0.47\textwidth}
      \includegraphics[width=\linewidth]{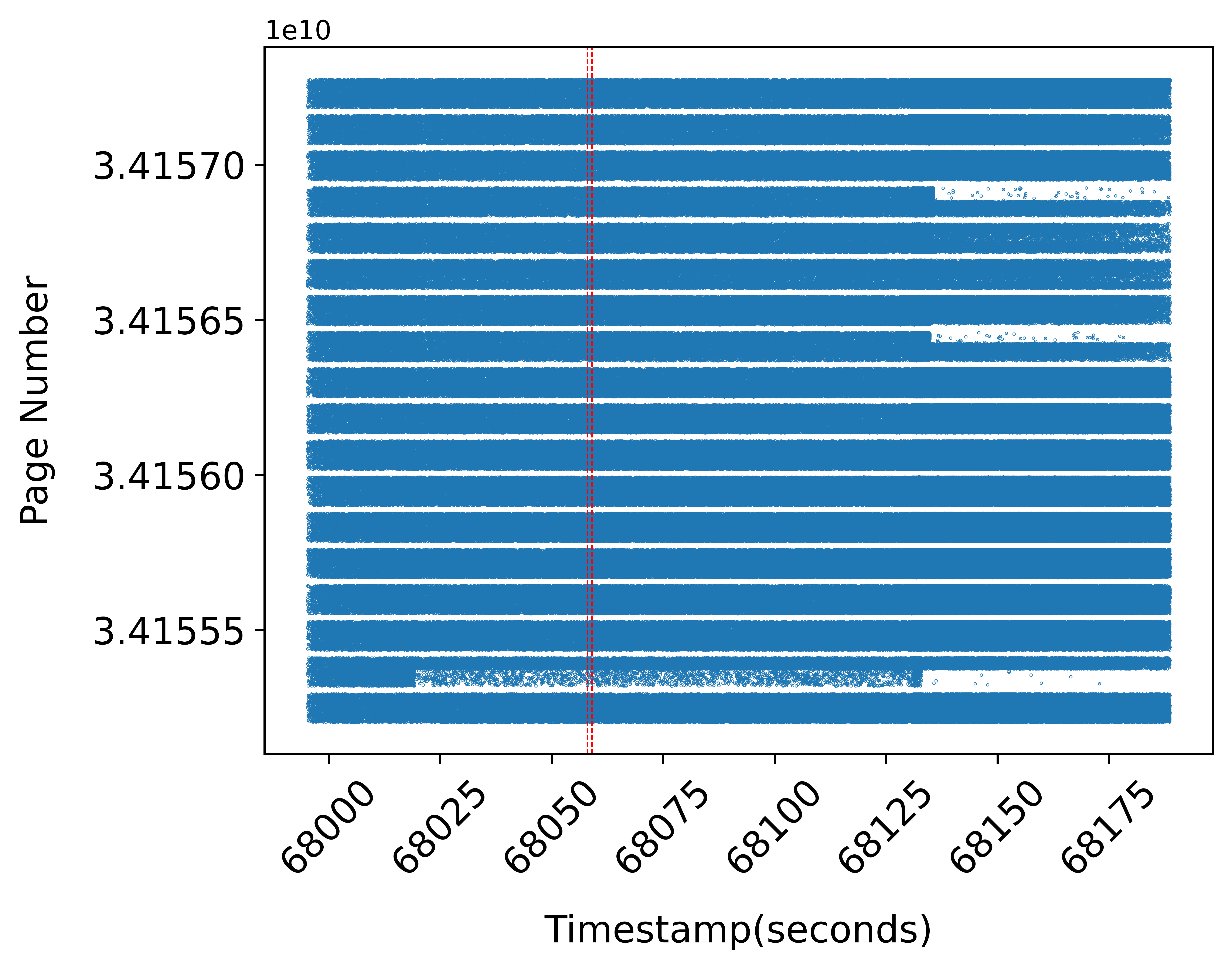}
      \caption{Entire Workload execution}
      \label{fig:1}
    \end{subfigure}\hfil 
    \begin{subfigure}{0.47\textwidth}
      \includegraphics[width=\linewidth]{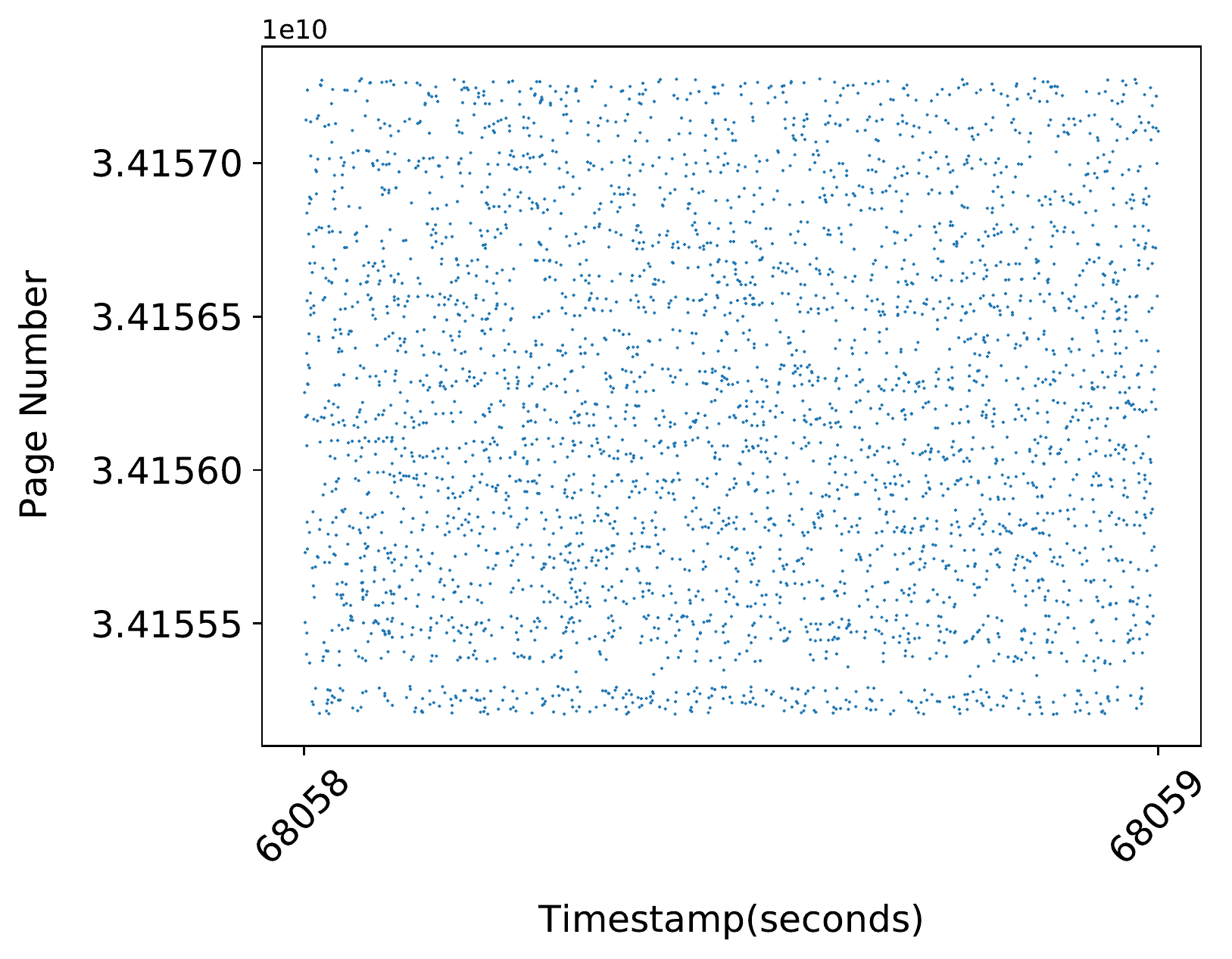}
      \caption{Zoom into one second interval}
      \label{fig:2}
    \end{subfigure}\hfil 
    \caption{Access pattern for the most accessed object pages in NVM over all time and over 1 second (using workload \emph{bc\_kron}). (b) refers to a sub-range of (a), where we can more clearly visualize the randomization of accesses.}
    \label{fig:access_pattern_top1_NVM}
\end{figure}

Figure \ref{fig:access_pattern_top1_NVM}(a) shows the access pattern throughout its lifetime for the object with the most access in the NVM (recall that object 0 has more than 60\% of accesses in NVM). On the x-axis we have the timestamp and on the y-axis the number of the page accessed. As the application has 18 threads, this object is accessed in parallel, with each thread accessing a different region of the object. At first, we may find that the access is easy to predict, but this happens due to the plot granularity. When we zoom in (1 second highlighted by red dashed lines in Figure \ref{fig:access_pattern_top1_NVM}(a) into Figure \ref{fig:access_pattern_top1_NVM}(b)), we realize that the access is actually random. In a random access pattern it is difficult to predict which pages will be accessed next.

\begin{tcolorbox}
\textbf{Finding 4}: The hottest objects of the application have random accesses and therefore their pages cannot be characterized as hot.
\end{tcolorbox}

\subsection{Tracking Memory Usage between NVM and DRAM}
\label{subsec:Tracking_Page_Distribution}

\begin{figure}[h]
    \centering
    \includegraphics[scale=0.87]{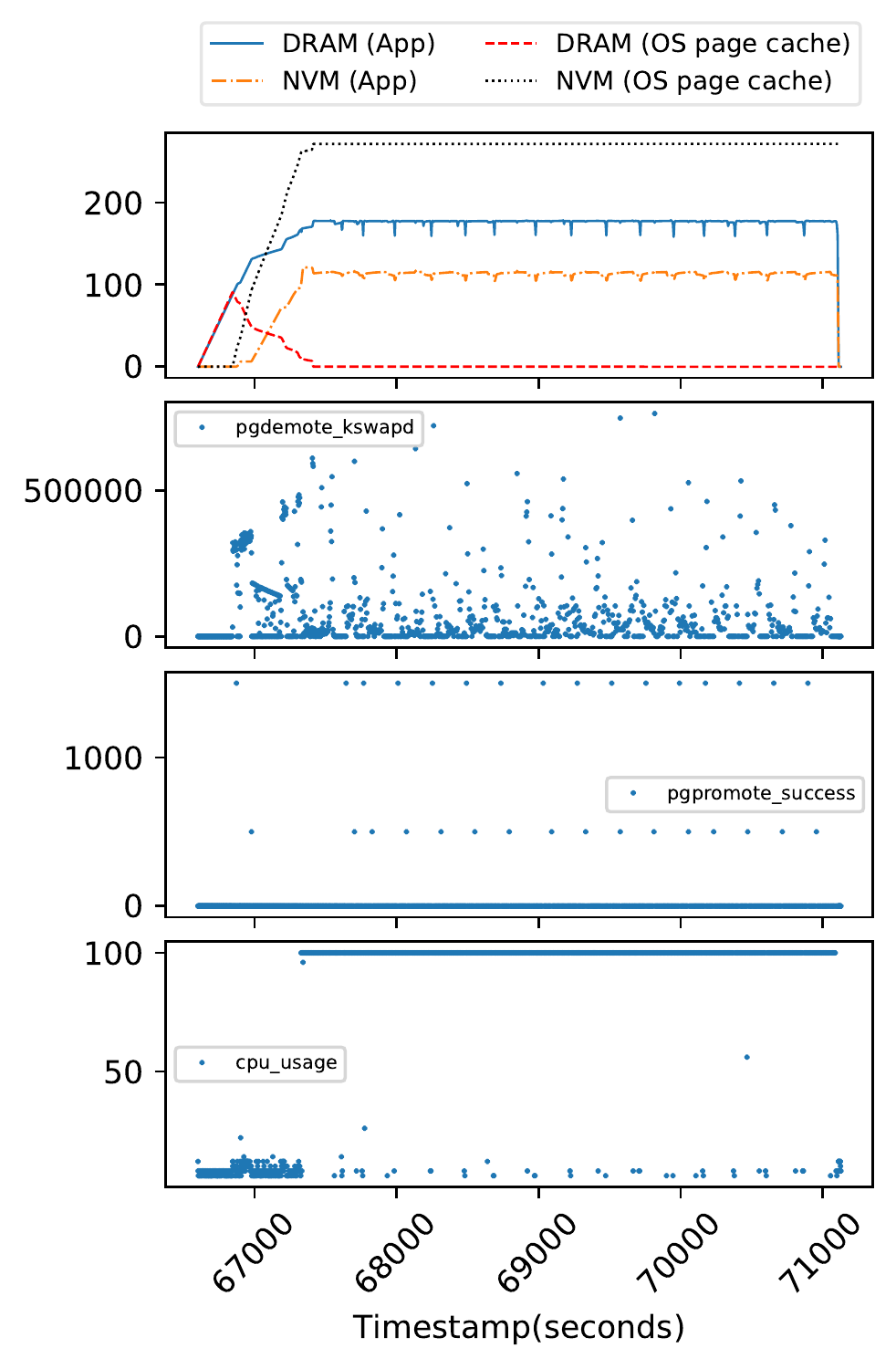}
    \caption{Memory usage in GB of workload \emph{bc\_kron} (top), demotion and promotion counters in number of pages (middle), and percentage of CPU usage (bottom) over time.}
    \label{fig:memory_cpu_counters}
\end{figure}

Figure \ref{fig:memory_cpu_counters} top graph shows the amount of memory allocated in DRAM and in NVM to \verb@bc_kron@ extracted from \verb@numastat@ Linux command. The other three graphs (top to bottom) show page demotion and promotion counters as well as CPU utilization.

Initially every allocation starts in DRAM (blue line). We noticed that when the application reaches approximately 100GB of data allocated in the DRAM, the new pages are allocated in the NVM. At first glance, this does not make sense, since the DRAM has 192GB. Of course, a small part of this is destined for the operating system itself. The issue is with the Linux \textit{Page Cache}: to optimize performance, whenever Linux does an I/O operation and there is free space in the DRAM, the page cache is activated to try to store data and avoid new I/O operations as much as possible. After the space in the DRAM is exhausted by the application itself and by the cache page, AutoNUMA starts its demotion process. Thereafter, any new allocations from the application and the page cache system happen in the NVM. The only other way to reduce DRAM allocations besides the operating system reclaim operation is the page demotion done by AutoNUMA. We can see (middle two plots of Figure \ref{fig:memory_cpu_counters}) that more demotions are performed compared to promotions. Furthermore, we can see that the CPU usage (bottom plot) is small at the beginning of execution due to the input reading phase of the graph. However, when the reading phase ends, CPU usage rises, coinciding with the end of new page cache allocations (see plot at the top). 
Page cache data was extracted from the Linux \verb@free@ command.

\begin{tcolorbox}
\textbf{Finding 5}: AutoNUMA was able to cut half of the memory capacity allocated by the OS for page cache through its demotion algorithm, while freeing space on DRAM for new allocations.
\end{tcolorbox}


\subsection{AutoNUMA Counters}

Page migrations between DRAM and NVM must be judiciously carried out to avoid unnecessary migration costs. To analyze the migrations that occur between NVM and DRAM and vice versa, we monitor counters like \verb@pgmigrate_success@ that counts page migration inter-socket. However, new counters (intra-socket) have been added to \verb@vmstat@ Linux command for heterogeneous memory context. We collected two  counters related to page promotion (NVM $\rightarrow$ DRAM). The first is \verb@pgpromote_success@, which is responsible for counting the number of pages successfully promoted. The second is \verb@pgpromote_demoted@, which only counts pages that are promoted and then demoted. It is important to analyze if there is thrashing between NVM and DRAM. Another two counters collected are related to page demotion (DRAM $\rightarrow$ NVM). The concept of demotion is similar to the well-known concept of reclaim. Reclaim operations are performed in two ways: periodic and direct. The first counter, \verb@pgdemote_kswapd@, counts the periodic reclaim performed by the kernel. The second counter, \verb@pgdemote_direct@, is more expensive and synchronous, counts the direct reclaim and occurs when a page allocation is requested but there is no more space in DRAM.

The values of these counters are cumulative since boot and we did not find information on how to reset them. Therefore, all values shown in this section are calculated through the difference (we call it \textit{delta}) of the respective counter at two different moments. To make sure the counters were working, we collected all four counters with AutoNUMA disabled. In this configuration, the pages allocated in the DRAM and in the NVM continue until the end of the execution, that is, there is no migration. All counters had zero delta. 

\begin{tcolorbox}
\textbf{Finding 6}: The number of promotions was found to be small and below to the rate limit as specified by AutoNUMA. This is because graph applications have very low page reuse over time, which makes it hard for AutoNUMA to identify promotion candidates.
\end{tcolorbox}

\subsection{Effectiveness of Promotions}

\begin{figure}[t]
    \centering
    \includegraphics[scale=0.6]{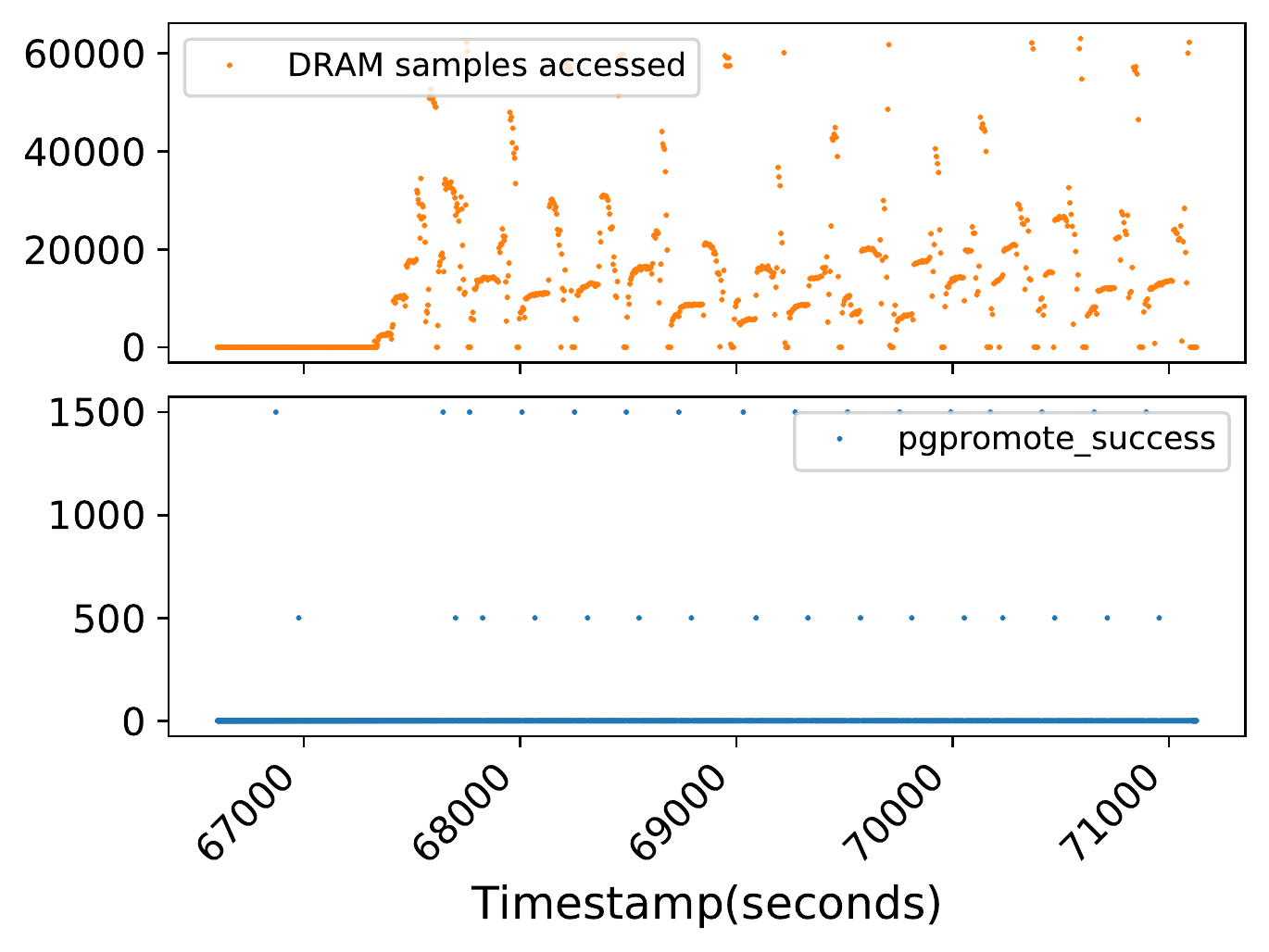}
    \caption{Number of load access in DRAM over time vs number of pages promoted to DRAM.}
    \label{fig:promoted_vs_dram_access}
\end{figure}

To verify the effectiveness of promotion-migrations, we analyzed the correlation between the number of promotions (\ref{fig:promoted_vs_dram_access} bottom) and the number of samples accessed in the DRAM (\ref{fig:promoted_vs_dram_access} top) over time. 
We notice that few pages are promoted over time. Furthermore, the maximum amount of promoted pages was much lower than the rate limit set by default by AutoNUMA (up to 65,536 Mbps, or 8GBps). Pages allocated in DRAM by AutoNUMA can come in two ways. The first case is when there is space in the DRAM. The second case is from hot pages in the NVM promoted to DRAM. This plot reinforces our hypothesis that most of the pages allocated in DRAM came from initial allocations.

\begin{tcolorbox}
\textbf{Finding 7}: Little correlation exists between number of promoted pages and number of pages accessed in DRAM. 
One hypothesis is that a large group of those pages that is considered hot by AutoNUMA (and therefore promoted) may belong to the group of pages with just two touches.
\end{tcolorbox}

\section{Opportunity for Object-level Allocation}

We briefly experiment with an idea for memory tiering based on object-level rather than page-level (as in AutoNUMA). This idea was derived after understanding the distinct characteristics of graph applications and examining the limitations of AutoNUMA in recognizing hot pages for these applications. 
First, we profile and perform a ranking of the application objects given by total memory accesses divided by allocation size (from high to low). Next, we assign objects to DRAM, starting from highest to lowest ranking, until the DRAM capacity is reached. Objects that cannot fit on DRAM are assigned entirely to NVM. The profiling and assignment steps are done once for each workload. After deciding (during the assignment step) which objects are assigned to each memory type, we use a shared library (\verb@syscall_intercept@) to keep track of each object allocation and map each object (memory region) to preassigned memory type (using Linux \verb@mbind@ system call). Once the binding policy of a memory region is altered, all pages in this range are allocated and remain in each type of memory until the end of the execution, thus no demotions or promotions are performed.

\begin{figure}[t]
    \centering
    \includegraphics[scale=0.99]{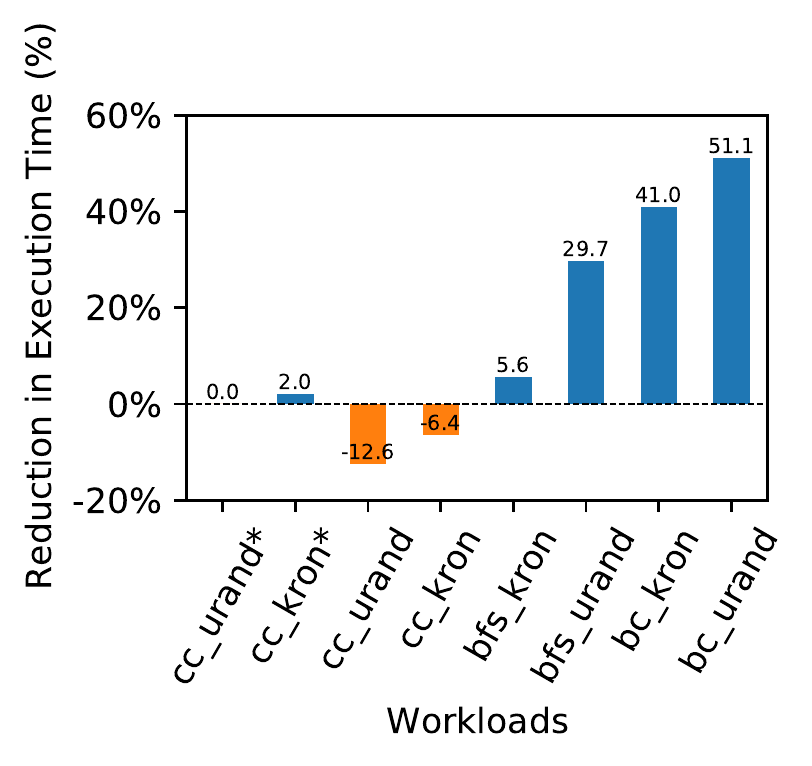}
    \caption{Performance improvements of object-level static mapping over AutoNUMA. Workloads \emph{cc\_urand*} and \emph{cc\_kron*} were executed using a variant of our proposal that spills over the object from DRAM to NVM.}
    \label{fig:gain_or_lost_exec_time}
\end{figure}

Figure \ref{fig:gain_or_lost_exec_time} summarizes the performance results of using this object-level approach compared to using AutoNUMA. We observed up to 51\% reduction of execution time over AutoNUMA. The performance results can be explained by how much memory access took place on NVM. For the \verb@bc_kron@ workload, as studied in detail in Section \ref{sec:AutoNUMA_track_decisions}, the object-level approach using static mapping reduced the number of memory samples taking place in NVM by 79\% compared to AutoNUMA. This, in turn, translated into 41\% faster execution time. 

It is worth mentioning that when an object does not fit entirely in the DRAM, it will be allocated to NVM. However, in case this a hot object (highly accessed), the application's performance will be negatively affected. Looking at Figure \ref{fig:gain_or_lost_exec_time}, this was particulaly unfavorable for workloads \verb@cc_urand@ and \verb@cc_kron@. Both workloads had performance slowdowns compared to AutoNUMA because of the way we assigned objects to DRAM; that is, either we assigned the entire object or no assignment at all. As we observed in Section~\ref{sec:AutoNUMA_track_decisions}, this could be even more pronounced if this object had a high number of TLB misses. This is because when the address translation is not found in caches, the page walk is done externally, which increases the latency of the memory operation. 


For workloads \verb@cc_urand@ and \verb@cc_kron@ showing performance slowdowns compared to AutoNUMA, we tried a variant of our proposal that spills over the object from DRAM to NVM. This variant is expected to improve performance or at least maintain the performance results obtained. The variant runs are shown as \verb@cc_urand*@ and \verb@cc_kron*@ in Figure \ref{fig:gain_or_lost_exec_time}. We observe that the performance of our proposal compared to AutoNUMA was the same for \verb@cc_rand*@ and an improvement of 2\% on workload \verb@cc_kron@ over AutoNUMA, whereas previously we had a slowdown of 6\% without performing the DRAM spilling. The difference between a workload with and without the asterisk (Figure \ref{fig:gain_or_lost_exec_time}) is how the static scheduler works. In workloads without an asterisk, all objects are allocated without any splitting, that is, they are either entirely allocated in DRAM or NVM. However, this increases the chances of leaving the DRAM capacity unused especially when you have large objects. On the other hand, in asterisked workloads, a single object is allowed to be spilled over NVM, that is, a part of it is placed in DRAM and another part in NVM, to improve DRAM utilization.



\section{Related Work}

The trend of heterogeneity goes beyond processors and is also consolidating with memory. 
In a context where there is a low-capacity and high-speed memory combined with a high-capacity and  lower-performance memory, new memory management solutions are required to extract the best balanced performance from each memory.
Several works have been proposed to scale applications in a heterogeneous memory context \cite{Dulloor2016,Effler19,chen2020,Malicevic2015,Olson2018,Servat2017,Wen2018}, none of which use Intel Optane technologies. \cite{Kannan2021, Ren2021 , Kumar2021, Raybuck2021} make use of Intel Optane however none compares with AutoNUMA. To the best of our knowledge, only two works \cite{Jonghyeon2021,Miguel2021} made a comparison with AutoNUMA. Kim et al. \cite{Jonghyeon2021} proposed the \textit{AutoTiering} algorithm as an alternative to AutoNUMA in which it had superior performance for most of the analyzed benchmarks. One of the few benchmarks in which \textit{AutoTiering} was not superior to AutoNUMA was graph500, precisely the type of application analyzed in our study. However, the authors did not elaborate much on why this happened. Only that the algorithm proposed by them resulted in more accesses to pages in NVM than AutoNUMA. Marques et al. \cite{Miguel2021} proposed the \textit{HyPlacer} algorithm and evaluated using the NAS Parallel benchmark rather than graph applications. Both \cite{Jonghyeon2021} and \cite{Miguel2021} make use of Intel Optane and compare their proposals with AutoNUMA. However, this comparison was made by analyzing only the execution time. It would be interesting to know not only that one approach was better or worse than the AutoNUMA, but also to understand why. Our work sought to understand by quantifying the decisions of the AutoNUMA and highlighting some limitations in its approach. 

\section{Conclusion}

In this paper, we characterize the performance of AutoNUMA for memory tiering using different graph applications on a real environment with DRAM and NVM. We study the memory management decisions by AutoNUMA and highlight its limitations on graph applications. To overcome the shortcomings of AutoNUMA, we propose a new idea based on  object-level (rather than page-level) memory tiering mapping. The idea works by intercepting the objects and deciding whether to allocate each one to DRAM or NVM. We compare our proposal of transparent offline object-level memory management with AutoNUMA. Our preliminary results show a performance improvements of up to 51\% in the execution time of graph applications, resulting from significantly minimizing memory accesses on the NVM.

%

\section*{Acknowledgments}
This work was partially funded by the Coordenação de Aperfeiçoamento de Pessoal Nível Superior - Brasil (CAPES) - Grant number 001. We would also like to thank the support of UFBA and IFBA, as well as the University of Pittsburgh, which provided access to the machine with Intel Optane.

\bibliographystyle{IEEEtranS}
\bibliography{reference}

%
%
%
%
%






\appendix
\section{Artifact Appendix}

\subsection{Abstract}

This document describes the process of acquiring and reproducing the experimental results presented in ``Performance Characterization of AutoNUMA Memory Tiering on Graph Analytics". This paper was accepted for publication at 2022 IEEE International Symposium on Workload Characterization (IISWC 2022). 

\subsection{Artifact check-list (meta-information)}

{\small
\begin{itemize}
  \item {\bf Algorithm: } Hottest object sorting
  \item {\bf Program: } Graph applications
  \item {\bf Compilation: } GCC compiler 
  \item {\bf Data set: } Kron and urand
  \item {\bf Run-time environment: } Any linux distribution with a kernel higher than version 5.1. The following tools installed: ipmctl, ndctl, daxctl and perf. Library Intercept Syscal must also be installed.
  \item {\bf Hardware: } CPU supporting Intel Optane, Intel Optane memories (At least 128GB) and 192 GB of DRAM
  \item {\bf Execution: } Bash and python scripts for automatic execution
  \item {\bf Metrics: } Execution time
  \item {\bf Output: } Results in CSV files and graphs
  \item {\bf Experiments: } A set of scripts will guide the experiment workflow.
  \item {\bf How much disk space required ?: } $\approx$470GB (datasets) + $\approx$160GB (traces)
  \item {\bf How much time is needed to prepare workflow ?: } $\approx$2h
  \item {\bf How much time is needed to complete experiments ?: } $\approx$16h
  \item {\bf Publicly available?: } Yes
  \item {\bf Archived (provide DOI)?: } 10.5281/zenodo.7058117
\end{itemize}
}

\subsection{Description}

\subsubsection{How to access}

The artifact code base can be obtained by downloading from \url{https://zenodo.org/record/7058117}. A more updated version is maintained on GitHub at \url{https://github.com/diegobnh/iiswc_22.git} .

\subsubsection{Hardware dependencies}

The CPU must support Intel Optane memory. We also need Intel Optane memory with at least 128GB of capacity. Finally, 192GB of DRAM memory is required in a given socket. 

\subsubsection{Software dependencies}

The kernel must be greater than or equal to kernel 5.1 to support the volatile-use of persistent memory as a hotplugged memory region (KMEM DAX). When this feature is enabled, persistent memory is seen as a separate NUMA node.
It is necessary to install the \verb@syscall-intercept@ library responsible for intercepting memory allocations in user space.
We also use the \verb@ndctl@ and \verb@ipmctl@ utilities to configure NVM and instantiate as NUMA node. Finally we need to have the \verb@perf@ software suite installed and enabled, which alredy happens in most Linux distributions.

\subsubsection{Data sets}

We use applications and datasets extracted from \verb@gapbs@. We selected from the GAPBS suite exclusively \verb@urand@ and \verb@kron@ inputs because the other datasets (\verb@twitter@, \verb@road@, and \verb@web@) had memory footprints too small to be used in our system. 

\subsection{Installation}
\label{installation}
\textbf{Cloning the Artifact Repository}. We have created a github repository to clone our artifacts: \url{https://github.com/diegobnh/iiswc\_2022}.

\textbf{Cloning the Library Intercept Syscall}. We have created a github repository for installing the library responsible for intercepting application allocations: \url{https://github.com/pmem/syscall\_intercept}. Perform the installation as described in the README file.

\textbf{Cloning the Workloads}. We have created a github repository for installing applications and datasets: \url{https://github.com/sbeamer/gapbs}. We changed the \verb@Makefile@ to compile with the -g flag and commented out the original outputs for all applications. Perform the installation as described in the README file. Then just compile the applications using the \verb@make@ command and finally generate the datasets with the commands:
\begin{itemize}
    \item ./converter -g30 -k16 -b kron.sg 
    \item ./converter -u31 -k16 -b urand.sg 
\end{itemize}

\textbf{Installing and Configuring Optane Memory.} Two utilities are used to configure memory: \verb@ipmctl@, \verb@ndctl@ and \verb@daxctl@ . The commands below are for the Fedora distribution, but similar commands can be run on other distributions.

\begin{itemize}
    \item sudo yum install ipmctl ndctl daxctl
\end{itemize}

The commands below, run with root permission, configure the Intel Optane as App Direct Mode on a single socket and then instantiate as on a node:

\begin{itemize}
    \item ipmctl create -f -goal -socket 0 PersistentMemoryType=AppDirect
    \item ndctl create-namespace --region=region0 -m devdax --map=mem
    \item daxctl reconfigure-device dax0.0 --mode=system-ram
\end{itemize}

\subsection{Experiment workflow}

After following the instructions in Section \ref{installation} above, clone the following repository: \url{https://github.com/diegobnh/iiswc_22.git}. Before running, the \verb@GAPBS_PATH@ environment variable must be setup with the path where the application was installed. Example: \verb@export GAPBS_PATH=/my_path@. Then you will enter the \verb@collect_scripts@ folder and run the following commands:
\begin{itemize}
    \item sudo ./start\_run.sh 
    \item sudo ./start\_post\_process.sh
    \item ./start\_mapping.sh
\end{itemize}

Once completed, enter the plot scripts folder and run the following commands:

\begin{itemize}
    \item ./generate\_inputs\_to\_plot.sh
    \item ./start\_plots.sh
\end{itemize}

\subsection{Evaluation and expected results}

Executing the \verb@start_run.sh@ script in folder \verb@collect_script@ will generate a folder with the respective workloads name (application\_dataset). Within each folder, there are the following folders: \verb@autonuma@ and \verb@static_mapping@ with the following files:

\begin{itemize}
    \item \verb@allocations.csv@
    \item \verb@track_info.csv@
    \item \verb@perf.data@
\end{itemize}

Each of these files has the workload name associated with the file name. After running the script \verb@post_process.sh@, three new files are generated in \verb@collect_script@ folder : 
\begin{itemize}
    \item \verb@memory_trace.csv@
    \item \verb@mmap_trace.csv@ 
    \item \verb@munmap_trace.csv@
\end{itemize}

Each of these files also has an associated workload name. The \verb@start_mapping.sh@ script will generate three new files:

\begin{itemize}
    \item \verb@mmap_trace_mapped.csv@
    \item \verb@perfmem_trace_mapped_DRAM.csv@
    \item \verb@perfmem_trace_mapped_PMEM.csv@
\end{itemize}

Once the generation and post processing phase is over, we run the scripts responsible for the article's graphics:

\begin{itemize}
    \item \verb@generate_inputs.sh@
    \item \verb@start_plots.sh@
\end{itemize}

\subsection{Experiment customization}

Benchmarks can be customized to create a smaller dataset. This will create shorter experiments. However, this is directly related to the RAM memory capacity. Our experiments were designed to ensure that the application's memory footprint does not fit into the DRAM.


\end{document}